\documentclass{article}

\usepackage{arxiv}
\usepackage{amsmath}
\usepackage[utf8]{inputenc} 
\usepackage[T1]{fontenc}    
\usepackage{hyperref}       
\usepackage{url}            
\usepackage{booktabs}       
\usepackage{amsfonts}       
\usepackage{nicefrac}       
\usepackage{microtype}      
\usepackage{cleveref}       
\usepackage{lipsum}         
\usepackage{graphicx}
\usepackage{natbib}
\usepackage{doi}
\usepackage{booktabs}
\usepackage{makecell}

\title{DetMesh-Gadep: Triangulated Surface Modeling and GPU-based Monte Carlo Efficiency Calibration of High-Purity Germanium Detectors}

\author{Kainan Zhang, Shuchang Yan, Zhen Wu, Hui Zhang, Rui Qiu and Junli Li}

\renewcommand{\shorttitle}{DetMesh-Gadep: Triangulated Surface Modeling and GPU-based Calibration}

\begin{document}
\maketitle

\begin{abstract}
Sourceless efficiency calibration of high-purity germanium (HPGe) detectors can provide accurate detector-response information without experiments using radioactive calibration sources, offering advantages in both convenience and safety. In many practical implementations, this process is performed using Monte Carlo simulation; however, its performance is constrained by the accuracy of detector modeling, the operational complexity of simulation frameworks, and the computational-resource requirements associated with CPU-based parallelization. In this study, a complete detector modeling and simulation framework is proposed. The detector modeling program DetMesh can generate triangulated surface geometry from parameterized detector models, providing advantages in the representation of complex geometric boundaries. It incorporates standard geometric operations and a geometric library, and is lightweight with strong extensibility. The generated geometry is then input into Gadep, a GPU-based Monte Carlo computational kernel, to enable rapid simulation. For $1\times 10^8$ particles, a single RTX 4090 achieved a speedup factor of 13.53 compared with simultaneous computation using 60 CPU cores. The proposed framework has low implementation cost and broad applicability, providing a complete solution for refined detector modeling and calibration.
\end{abstract}


\section{Introduction}\label{sec1}

High-purity germanium (HPGe) detectors remain the reference standard for high-resolution gamma ray spectrometry because they provide an exceptional compromise between energy resolution and full-energy peak efficiency (FEPE) over a broad energy range \cite{001knoll2010radiation,002eberth2008ge,TSOU199430}. In quantitative gamma analysis, however, converting peak areas into activities or emission rates requires an accurate FEPE calibration for the exact detector–source configuration. This requirement becomes particularly demanding when samples differ from available standards in geometry, composition, or source-to-detector distance. When routine handling of radioactive calibration sources is impractical, sourceless efficiency calibration becomes a safe and operationally feasible alternative, and has therefore remained a longstanding focus in HPGe metrology \cite{006moens1981calculation,007sima2002transfer}.

Sourceless efficiency calibration is not tied to a single computational route. Analytical and semi-empirical approaches based on effective-solid-angle formulations and efficiency transfer remain attractive because they are fast and convenient once a reference calibration is available \cite{006moens1981calculation,b022moens1983peak,b023wang1995esolan,b024abbas2007direct}. Software implementations such as GESPECOR and ETNA further extend this strategy to matrix corrections, geometry transfer, and coincidence-summing treatment \cite{007sima2002transfer,b027radu2009etna}. Hybrid strategies have likewise been explored, either by combining simplified simulation with reference measurements or by coupling Monte Carlo output with semi-empirical transfer schemes to reduce calibration effort \cite{b025jiang1998hybrid,b026ozben2009hybrid,b028salman2019hybrid}. For routine sourceless calibration across diverse source matrices and non-reference geometries, however, full Monte Carlo transport remains the most general, flexible, and dependable framework, because it treats particle transport and detector geometry in a unified manner without requiring a new problem-specific derivation for each measurement configuration \cite{003agostinelli2003geant4,004allison2016recent,005briesmeister2000mcnptm}.

Landmark studies have shown that Monte Carlo simulations can reproduce HPGe efficiencies with high fidelity when the detector structure is described with sufficient realism \cite{008hurtado2004monte,009boson2008detailed,010budjavs2009optimisation}. For coaxial HPGe detectors, FEPE is sensitive to dead layers and crystal geometry, especially at low energies and in non-standard counting geometries \cite{011vidmar2009crystal,012subercaze2022effect}. In practice, therefore, the more persistent challenge lies not in the transport formalism alone, but in the accurate characterization of the detector and in the efficient coupling of detector geometry to transport calculation. This point has been repeatedly underscored in both semi-empirical and Monte Carlo studies, which show that reliable computed efficiencies depend critically on accurate information about detector construction, optimized geometric parameters, and dead-layer characterization \cite{b029bell2012angle,b030prozorova2021characterizing,b031lin2023determination}.

This challenge becomes more severe in sourceless efficiency calibration workflows, where multiple energies, geometries, or detector instances must be evaluated repeatedly. Although commercial and research software has greatly expanded the accessibility of model-based calibration, detailed detector descriptions still tend to rely on general-purpose constructive-solid-geometry representations or on proprietary detector characterizations. These approaches are effective, but they are not always easy to embed in streamlined, high-throughput computation. A compact geometric representation that preserves the physically consequential features of HPGe crystals—especially dead layers, internal bores, and rounded or bulletized front surfaces—would therefore remove an important bottleneck in routine sourceless efficiency calibration.

That flexibility, however, comes at a computational cost. Monte Carlo radiation transport has long served as a core computational framework in nuclear science and radiation measurement, and general-purpose platforms such as MCNP and Geant4 have accordingly become standard tools in detector-response simulation, shielding analysis, and related radiation-transport problems \cite{b017goorley2012initial,niess2025goupil,soplin2025monte}. Statistically reliable results usually require very large numbers of particle histories, and convergence can be especially slow in deep-penetration, geometry-rich, or detector-response calculations \cite{b019garcia2021variance,b020asano2022photon,b021zhang2023development}. In efficiency calibration studies, where detector responses may need to be evaluated repeatedly over multiple energies or measurement configurations, this cost readily becomes a practical challenge \cite{b020asano2022photon,b021zhang2023development}.

At the same time, the case for GPU acceleration is compelling. Monte Carlo transport is naturally parallel at the particle-history level, and GPU-based engines have already reshaped neighboring areas of radiation transport. Representative systems such as gDPM, GPUMCD, goMC, and FRED have demonstrated that carefully designed GPU algorithms can deliver substantial speedups while preserving agreement with established Monte Carlo benchmarks \cite{013jia2011gpu,014hissoiny2011gpumcd,015tian2015gpu,016franciosini2023gpu,ren2026research}. Yet these advances have been driven mainly by dose calculation in voxelized or otherwise geometries. Detector-efficiency calibration poses a different challenge: the transport engine must be coupled to detector-specific geometry with sufficient fidelity to represent coaxial HPGe structures, dead layers, bulletized crystal ends, and internal bores.

In this work, we bridge these two tracks by coupling a lightweight faceted-geometry framework, DetMesh, with the GPU-accelerated Monte Carlo engine Gadep. DetMesh converts detector parameters directly into triangulated surface models that preserve the structural features most relevant to sourceless efficiency calibration, while Gadep performs photon–electron transport on the GPU using these meshes as native geometric input. The resulting workflow is designed for sourceless efficiency calibration of HPGe detectors under practical assay conditions, where both geometric fidelity and computational throughput are decisive. By unifying parameterized detector modeling and GPU transport in a single pipeline, this study provides a scalable alternative to conventional CPU-centered calibration workflows and extends high-accuracy Monte Carlo efficiency calibration toward routine, rapid deployment.

\section{Materials and Methods}\label{sec2}

\subsection{Detector instances}

To validate and evaluate the performance of the detector modeling and accelerated Monte Carlo simulation program proposed in this study for practical sourceless efficiency calibration, six coaxial HPGe detectors manufactured by ORTEC were selected as experimental validation objects.

The selected detector set covers different models and crystal structures, mainly including p-type coaxial detectors from the GEM series and n-type coaxial detectors from the GMX series. The geometric parameters of all detectors were directly extracted from the quality assurance documents provided by the manufacturer. These nominal parameters constitute the physical basis for generating three-dimensional triangulated surface models using the DetMesh framework. The detailed geometric parameters are listed in Table \ref{tbl1}.

\begin{table}
\caption{The detector instances modeled and simulated in this study}\label{tbl1}
\begin{tabular}{@{}llccccccc@{}}
\toprule
\textbf{No.} & \textbf{Model}  & \textbf{Type} & \makecell[c]{\textbf{Crystal}\\\textbf{Diam. (mm)}} & \makecell[c]{\textbf{Crystal}\\\textbf{Length (mm)}} & \makecell[c]{\textbf{Hole}\\\textbf{Diam. (mm)}} & \makecell[c]{\textbf{Hole}\\\textbf{Depth (mm)}} & \makecell[c]{\textbf{Dead Layer (mm)}\\\textbf{(Outer / Inner)}} \\ \midrule
\#1 & GMX45P4-76    & N-Type & 61.8 & 61.9 & 8.3  & 60.0 & 0.0003 / 0.7000 \\
\#2 & GEM-C50-LB-C     & P-Type & 68.1 & 63.3 & 10.1 & 50.2 & 0.7000 / 0.0003 \\
\#3 & GEM-C50P4        & P-Type & 67.9 & 65.6 & 8.5  & 52.5 & 0.7000 / 0.0003 \\
\#4 & GEM-C50P4        & P-Type & 68.7 & 65.6 & 8.6  & 52.0 & 0.7000 / 0.0003 \\
\#5 & GEM-C50P4        & P-Type & 68.5 & 65.0 & 8.7  & 52.2 & 0.7000 / 0.0003 \\
\#6 & GMX50P4-83       & N-Type & 68.8 & 65.8 & 8.6  & 52.2 & 0.0003 / 0.7000 \\ \bottomrule
\end{tabular}%
\end{table}

The data show that detectors from different series exhibit significant structural differences in the distribution of the dead layer. GMX-series detectors \#1 and \#6 exhibit typical structural characteristics of n-type detectors, with an extremely thin boron ion-implanted layer (0.3$\mu$m) on the outer surface, making them suitable for low-energy photon measurements. In contrast, GEM-series detectors \#2–\#5 have a thicker outer dead layer (0.7mm). By parsing these structural parameters, the DetMesh framework enables accurate detector modeling and thereby allows accurate calculation of energy deposition in subsequent particle transport simulations.

Furthermore, comparison of detectors \#3, \#4, and \#5, which are of the same model type, GEM-C50P4, shows that non-negligible individual differences remain in crystal diameter (67.9-68.7mm) and borehole length (52.0-52.5mm), even for detectors of the same model. These differences arise from crystal-growth processes and machining tolerances, and they can affect the detection efficiency of the detector. This further emphasizes the importance of accurate detector modeling and sourceless efficiency calibration.

\subsection{General modeling framework of DetMesh}

Radiation detectors are relatively precise and complex physical instruments. However, the geometric models corresponding to their main components are often not complicated. For example, the germanium crystal in some HPGe detectors can be represented as a cylinder with a front-edge rounding, while the crystal in a scintillation detector can be represented as a cuboid. General-purpose mesh generation software is often designed for objects with substantially higher modeling complexity, such as human bodies, buildings, and large mechanical facilities. Although such software provides a wide range of functions, part of its computational capability is redundant for detector modeling scenarios, where the required geometric processing is comparatively simple. General-purpose Monte Carlo transport codes also provide geometric modeling capabilities; however, their geometry modules are strongly coupled with the Monte Carlo computational kernels. In addition, their deployment cost is relatively high, making them difficult for ordinary users to implement in practice.

At the same time, detector modeling is characterized by the repeated invocation and assembly of simple geometric resources, which provides a natural basis for parameterized detector modeling. To provide a detector modeling solution that accurately meets this requirement, this study proposes the DetMesh framework. DetMesh adopts a hierarchical architecture that decouples geometric definition from physical assembly, allowing users to extend parameterized triangulated surface models for different detectors by defining basic primitives and assembly logic.

\begin{figure}
  \centering
   \includegraphics[width=.9\textwidth]{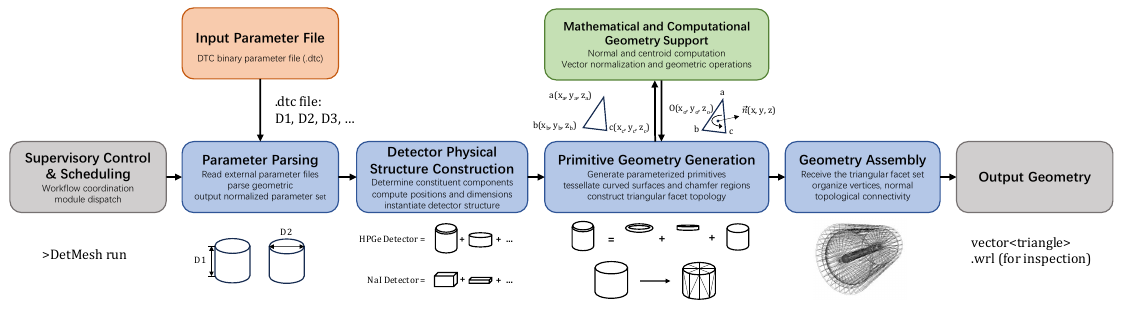}
    \caption{Framework of DetMesh, a general-purpose detector triangulated surface modeling software}\label{fig1}
\end{figure}

The functions and workflow of each program module are shown in Figure \ref{fig1}.

\subsubsection{Interfaces and architecture}

\begin{figure}
  \centering
   \includegraphics[width=1\textwidth]{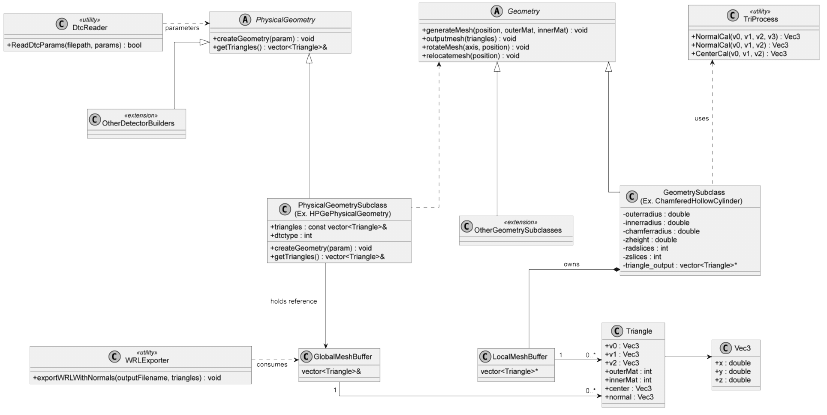}
    \caption{UML diagram of the DetMesh code structure}\label{fig2}
\end{figure}

The core of DetMesh achieves high extensibility through abstract base classes in C++. Figure \ref{fig2} presents the code structure of DetMesh using the Unified Modeling Language (UML).

For physical modeling, the framework defines the PhysicalGeometry interface, which contains the pure virtual function createGeometry. Any specific detector type can be incorporated into the framework by inheriting from this class and overriding the corresponding construction logic. This polymorphic design ensures that the system can accept standardized parameter-array inputs without modifying the underlying driver code.

At the geometric layer, the generation of low-level geometric models relies on the Geometry interface. The program uses a unified parent class, Geometry, to manage the triangulated-surface generation interface. This design standardizes mesh generation, spatial transformation, and data output methods for the geometric model library, while ensuring the extensibility of the library.

\subsubsection{Discretization and normal calculation}

The framework incorporates triangulated-surface discretization algorithms to convert continuous mathematical geometries into discrete surface data. It also provides discretization-accuracy parameters, allowing users to select appropriate triangulated surface models for different scenarios. For geometries with symmetry, such as cylinders and spheres, DetMesh adopts an angularly equidistant sampling method, which is well suited for features such as the front-edge rounding of a cylinder.

For example, for a cylinder rounded at one end, this strategy applies angular subdivision both in the radial direction and along the polar-angle direction of the rounded region. This approach ensures a smooth transition of curvature features while avoiding repeated node solving under equal spacing along the $Z$ direction. As a result, a finer and more reliable triangulated surface model can be constructed with reduced surface-generation time.

For the main body of the detector, namely the cylindrical surface, the algorithm performs angularly equidistant sampling on the $X-Y$ plane based on the radial discretization parameter $N_\theta$ For any point on the cylinder, the vertex coordinates $\left( x_i, y_i \right)$ are determined by the discrete angle $\theta_i$:

\begin{align}
    \theta_i = \frac{2\pi \cdot i}{N_{\theta}}, \quad i \in [0, N_{\theta}-1]
\end{align}

\begin{align}
    \begin{cases}
    x_i = R \cdot \cos\theta_i \\
    y_i = R \cdot \sin\theta_i
    \end{cases}
\end{align}

The front end of an HPGe detector is usually bulletized, meaning that a rounded region with radius $R_{\text{chamfer}}$ exists at the top of the cylinder. To accurately describe this modeling feature, DetMesh introduces the rounding subdivision parameter $N_\phi$:

\begin{align}
    \phi_j = \frac{\pi}{2} \cdot \frac{j}{N_{\phi}}, \quad j \in [0, N_{\phi}]
\end{align}

On this basis, the three-dimensional coordinates $\left( x_i, y_i, z_i \right)$of the vertices on the rounded surface are derived as:

\begin{align}
    \begin{cases}
    \rho_j = \left(R_{\text{out}} - R_{\text{chamfer}}\right) + R_{\text{chamfer}} \cdot \cos\phi_j \\
    z_j = Z_{\text{base}} + R_{\text{chamfer}} \cdot \left(1 - \sin\phi_j\right) \\
    x_{i,j} = \rho_j \cdot \cos\theta_i \\
    y_{i,j} = \rho_j \cdot \sin\theta_i
    \end{cases}
\end{align}

In Monte Carlo transport calculations, the normal direction of a triangular facet determines whether a particle is entering or leaving a medium. DetMesh integrates an independent mathematical calculation module, triprocess, which dynamically calculates the unit normal vector n of each triangular facet using the vector cross-product algorithm:

\begin{align}
    \mathbf{n} = \text{normalize}\left(\left(\mathbf{v}_1 - \mathbf{v}_0\right) \times \left(\mathbf{v}_2 - \mathbf{v}_0\right)\right)
\end{align}

This calculation is decoupled from specific geometric shapes, ensuring that every generated triangular facet carries correct directional information regardless of the complexity of the geometric model.

\subsubsection{Data structure and output}

DetMesh uses the Triangle structure as the minimum data unit. This structure encapsulates vertex coordinates, the normal vector, and the material attributes on the inner and outer sides of each triangular facet. This design enables the framework to handle interfaces between multiple materials. The final model is exported by the OutputGeometry module in the standard VRML 2.0 format. This format supports the indexed face set representation, which can be used in general-purpose 3D software for visual verification of the geometry.

In addition, Gadep is embedded as part of the program in this study. The subsequent modules or programs can directly invoke the in-memory vector data of triangular facets generated by DetMesh, thereby eliminating the need for file input and output and achieving higher data-transfer efficiency.

\subsection{Parametric modeling of HPGe Detectors}

This study models the single-ended coaxial HPGe detector instances described in the previous section. The modeling process mainly includes the physical mapping of the parameter space and the topological construction logic of key components.

\subsubsection{Geometric parameter space}

\begin{figure}
  \centering
   \includegraphics[width=.9\textwidth]{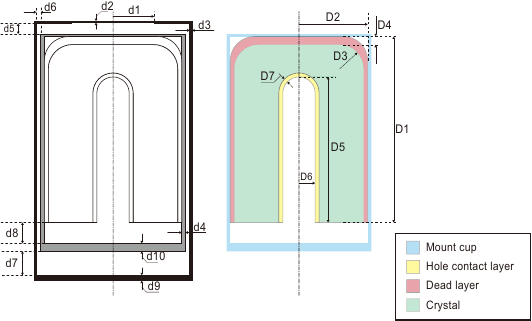}
    \caption{Geometric modeling parameters of a single-ended coaxial HPGe detector: the left panel shows the external structural parameter group, and the right panel shows the internal structural parameter group}\label{fig3}
\end{figure}

The geometric features of the detector are abstracted into a set of high-dimensional parameter vectors. According to the technical drawings and physical definitions provided by the manufacturer, the input parameters are divided into an external structural parameter group $\mathbf{P}_{ext}$ and an internal crystal parameter group $\mathbf{P}_{int}$, as shown in Table \ref{tbl2} and Figure \ref{fig3}.

\begin{table}
\caption{Geometric modeling parameters and definitions for a single-ended coaxial HPGe detector}\label{tbl2}
\centering
    \begin{tabular}{l l}
        \toprule
        \textbf{Parameter} & \textbf{Definition} \\
        \midrule
        d1     & entrance-window radius \\
        d2     & entrance-window thickness \\
        d3     & vacuum housing thickness \\
        d4     & crystal cup thickness \\
        d5    & top vacuum thickness \\
        d6    & side vacuum thickness \\
        d7    & negative-z height of the vacuum housing \\
        d8    & negative-z height of the crystal cup \\
        d9    & bottom thickness of the vacuum housing \\
        d10    & bottom thickness of the crystal cup \\
        D1   & crystal length \\
        D2   & crystal radius \\
        D3   & front-edge rounding radius \\
        D4   & dead-layer thickness \\
        D5   & borehole length \\
        D6   & borehole radius \\
        D7   & contact-layer thickness \\
        \bottomrule
    \end{tabular}
\end{table}

\subsubsection{Construction of key components}

Based on the geometry-generation algorithm of DetMesh, the construction of detector triangular facets follows the principles of inside-to-outside modeling and functional layering.

The main modeling challenges for a coaxial HPGe detector lie in the bulletized front-end shape and the geometric stripping of the outer dead layer. DetMesh uses the ChamferedHollowCylinder geometry to achieve accurate discretization of this structure through a component-wise construction strategy.

First, the physical outer contour of the crystal is constructed. When the triangulated-surface generation function is called, the zslices parameter is enabled to perform angularly equidistant sampling of the rounded region, generating a solid-boundary model with a smoothly transitional edge.

For the core sensitive region, a geometric shrinkage operation is performed during construction of the sensitive volume in order to physically distinguish the dead layer. The construction parameters of the sensitive region are dynamically modified as:

\begin{align}
    \begin{cases}
    R_{\text{active}} = D_2 - D_4  \\
    L_{\text{active}} = D_1 - D_4  \\
    \end{cases}
\end{align}

The generated mesh of the sensitive region is nested inside the outer shell. In Monte Carlo transport, the annular space between these two mesh surfaces is defined as the dead-layer region, thereby avoiding the stair-step errors associated with voxelized modeling.

Meanwhile, to accommodate structural differences among detector models, such as variations in cold-finger length, the modeling logic incorporates an adaptive assembly algorithm. The code calculates the vacuum distance behind the entrance window through logical judgment. When the distance between the crystal and the entrance window changes, the program automatically adjusts the height of the vacuum layer, ensuring that geometric overlap does not occur. For the coaxial borehole inside the crystal, the system constructs a rounded hollow cylinder using parameter $D_6$, the borehole radius, and $D_5$, the borehole length, and positions it according to parameter $d_8$, the negative-z height of the crystal cup.

Finally, all independently generated component meshes are mapped into the global coordinate system using the relocatemesh function. Taking the bottom of the vacuum housing as the reference datum $Z_{\text{offset}}$, the code sequentially stacks the electrode, crystal, crystal cup, vacuum housing, and other components along the z-axis, ultimately outputting complete triangular-facet data.

\subsection{GPU-based Monte carlo simulation}

The triangulated surface models generated by the DetMesh framework must ultimately be input into a Monte Carlo engine for particle transport simulation. In this study, the GPU-based coupled photon–electron transport program Gadep, namely the GPU-based Accelerated Dose Estimation Program, was adopted as the computational kernel \cite{YanGadep}. Gadep is developed based on the NVIDIA CUDA architecture and adopts a host–device collaborative parallel mode. Compared with traditional CPU-based general-purpose Monte Carlo programs, it achieves orders-of-magnitude speedup while maintaining computational accuracy.

\begin{figure}
  \centering
   \includegraphics[width=0.9\textwidth]{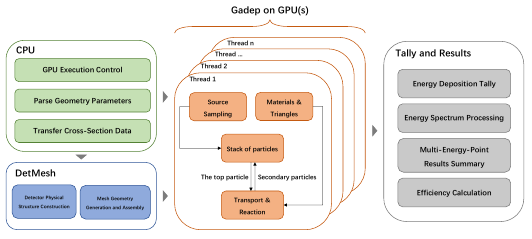}
    \caption{Coupling architecture of the Gadep–DetMesh program: the CPU controls DetMesh-based geometry generation and Gadep initialization, and performs result scoring after GPU Monte Carlo transport is completed}\label{fig4}
\end{figure}

As described above, to improve the reading speed of detector triangulated-surface data, the triangular-facet data output by DetMesh are stored in runtime memory and directly read by Gadep as program variables. This approach eliminates the transfer overhead caused by reading and writing triangulated-surface data files, forming a strongly coupled design between DetMesh and Gadep. It also further demonstrates the integrability and lightweight nature of DetMesh. The coupled workflow is shown in Figure \ref{fig4}.

\subsubsection{Physical models and parallel architecture}

Gadep simulates coupled photon–electron transport over an energy range from 100 eV to 100 GeV, satisfying the physical requirements of sourceless efficiency calibration for HPGe detectors. The program integrates the Livermore low-energy electromagnetic physics model to handle the photoelectric effect, Compton scattering, and pair production. It uses the Seltzer–Berger model to describe bremsstrahlung and adopts the Urban model to treat multiple scattering of charged particles.

Meanwhile, to adapt to the single-instruction multiple-thread architecture of GPUs, Gadep uses the partial cross-section method, rather than the traditional total cross-section method, for step-length sampling. This effectively reduces the frequency of random-number generation and optimizes the execution efficiency of GPU warps.

\subsubsection{Localization logic of particles}

In Gadep, the determination of the material in which a particle is located depends on a predefined topological nesting relationship, that is, the OuterMaterial must be explicitly specified during geometry definition. However, the DetMesh framework is designed to allow users to freely construct parameterized detector components. It cannot require users to know and define all complex geometric nesting hierarchies in advance, such as multilayer shielding structures or irregular cold-finger components.

\begin{figure}
  \centering
   \includegraphics[width=0.9\textwidth]{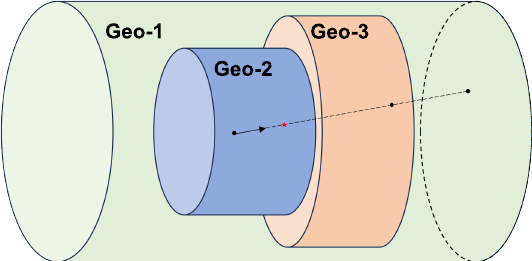}
    \caption{Modified geometry-identification method in Gadep for determining the geometry containing a particle. Geo-2 and Geo-3 share one coincident surface, and the red star denotes the intersection between the inner surface of Geo-2 and the ray along the particle flight direction}\label{fig5}
\end{figure}

For this reason, the underlying material-identification algorithm of Gadep was reconstructed in this study. In the original algorithm, a ray is emitted from the particle along its flight direction, and the nearest intersection with a triangular facet is identified. The normal direction of that facet is then used to determine whether the particle is entering or leaving the geometry. If the particle is leaving, the program directly reads the preset external material ID of that geometry. In the Gadep–DetMesh adapted version, the program no longer relies on preset external material information. When a particle undergoes cross-boundary transport, the kernel calculates the intersections between the particle ray and all geometries in the scene, or a subset filtered by bounding boxes, as shown in Figure \ref{fig5}.

The algorithm counts the number of intersections between the ray and each closed geometry. According to the three-dimensional extension of the Jordan curve theorem, if the number of intersections between the ray and a given geometry is odd, the particle is determined to be inside that geometry. Among all geometries satisfying the odd-intersection condition, the geometry whose first intersection is closest to the particle is selected as the container in which the particle is currently located. The internal material of this geometry is then assigned as the current environmental medium of the particle. If no geometry satisfies the odd-intersection condition, the particle is determined to be located in the world medium.

At the same time, when geometries of different materials, such as the dead layer, germanium crystal, borehole, and vacuum inside the borehole of an HPGe detector, have coincident bottom surfaces, the ray emitted from the particle along its flight direction may simultaneously hit the bottom surfaces of these geometries with different materials. This can make it impossible to determine the particle location unambiguously.

\begin{figure}
  \centering
   \includegraphics[width=0.9\textwidth]{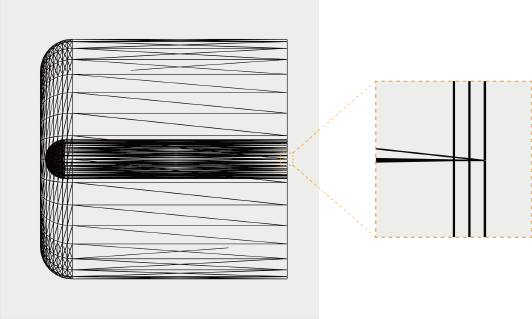}
    \caption{Treatment of internally coincident surfaces in nested geometries by DetMesh. In the enlarged view on the right, the surfaces from left to right correspond to the vacuum inside the borehole, the borehole, and the germanium crystal}\label{fig6}
\end{figure}

Therefore, during geometric assembly, these coincident bottom surfaces are separated algorithmically. This operation is performed from the inside outward. It ensures that the adjusted geometric-scale differences can be correctly recognized by the Monte Carlo transport module and that no geometric overlap is introduced, while avoiding any influence on the accuracy of Monte Carlo transport. Figure \ref{fig6} shows the modeling result of the assembled geometric modules and a local enlarged view.

\subsubsection{Acceleration Implementation}

Although the above parity-check-based identification logic has higher computational complexity for a single geometry query than the original table-lookup method, it removes the dependence of physical modeling on manually defined nesting relationships and enables plug-and-play use of the parameterized models generated by DetMesh.

Meanwhile, Gadep already uses a tree-based acceleration structure based on a bounding volume hierarchy (BVH) to store triangular-facet data. Even with the more complex geometric identification logic, the Gadep–DetMesh system still achieves orders-of-magnitude speedup over traditional MPI-parallel CPU Monte Carlo programs when simulating the full-energy peak efficiency of typical HPGe detectors. This performance fully satisfies the computational-efficiency requirements of rapid sourceless efficiency calibration.

\section{Results}\label{sec3}

The integrated framework is validated from three aspects: the accuracy of geometric modeling, the accuracy of sourceless efficiency calibration, and the computational performance of GPU acceleration. The six HPGe detectors described above were selected as test objects, with emphasis placed on the capability of the model to reproduce complex geometric features and on the efficiency advantage of the computational kernel when handling simulations with large numbers of particles.

\subsection{Modeling Results}

\begin{figure}
  \centering
   \includegraphics[width=.9\textwidth]{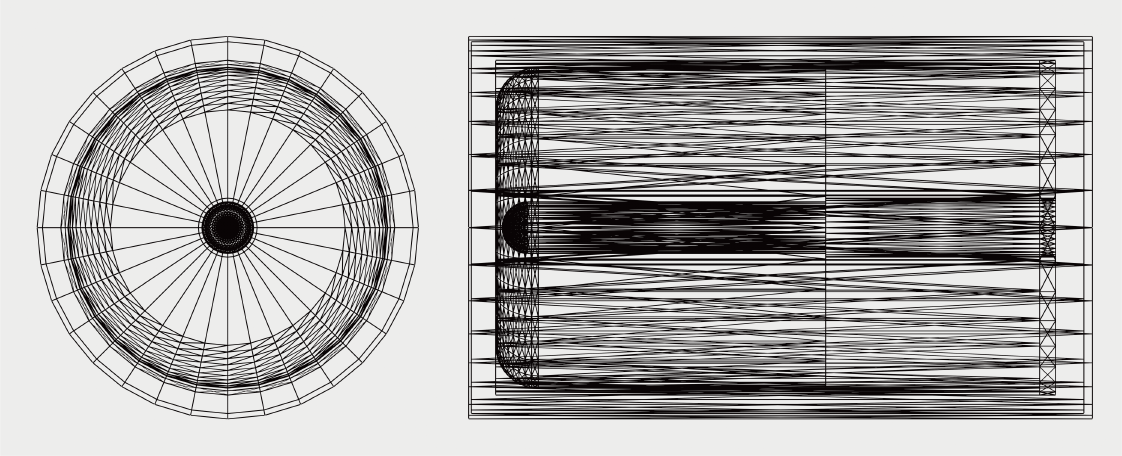}
    \caption{Axial and side views of the triangulated surface model of detector \#1}\label{fig7}
\end{figure}

\begin{figure}
  \centering
   \includegraphics[width=.9\textwidth]{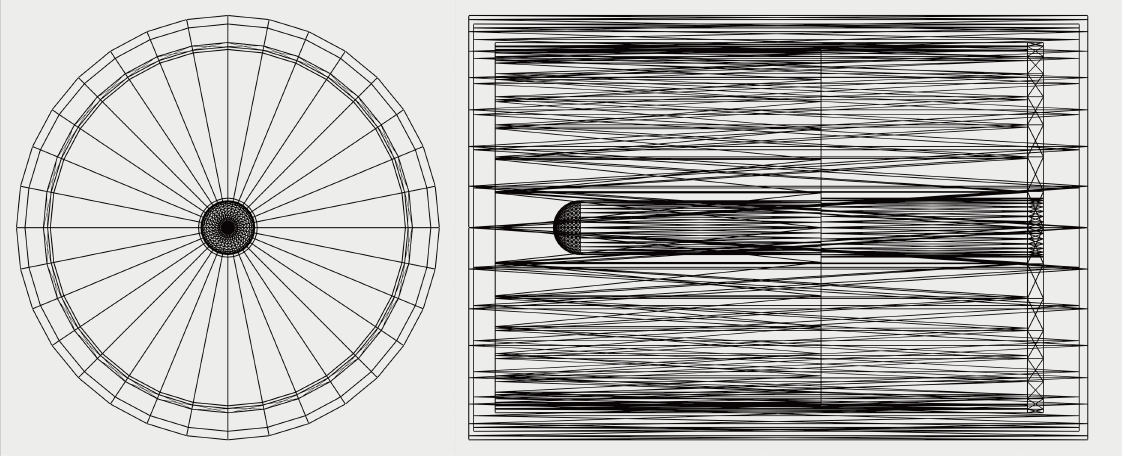}
    \caption{Axial and side views of the triangulated surface model of detector \#2}\label{fig8}
\end{figure}

Based on the DetMesh parameterized framework, three-dimensional triangulated surface models were successfully constructed for all detectors under investigation. Figure \ref{fig7} and \ref{fig8} shows the external modeling results of an n-type detector (\#1, GMX series) and a p-type detector (\#2, GEM series). It can be seen that DetMesh accurately reproduces the geometric differences among different detector models.

\begin{figure}
  \centering
   \includegraphics[width=0.9\textwidth]{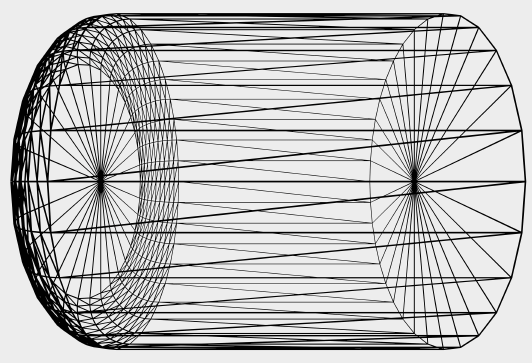}
    \caption{Side view of the germanium crystal component of detector \#1 with perspective rendering; the front-edge rounding structure of the germanium crystal is visible on the left}\label{fig9}
\end{figure}

To more clearly demonstrate the modeling capability of DetMesh for front-edge-rounded cylinders, several important components of detector \#1 were constructed separately, as shown in Figure \ref{fig9}. This figure shows the front-edge rounding region of the crystal. With an angular subdivision accuracy of $N_\phi = 10$, the generated mesh exhibits a smooth transition and clearly presents the double-layer nested structure in which the sensitive region is uniformly wrapped by the outer dead layer. This verifies the effectiveness of the component-wise construction strategy described in Section \ref{sec2}.

\begin{figure}
  \centering
   \includegraphics[width=0.9\textwidth]{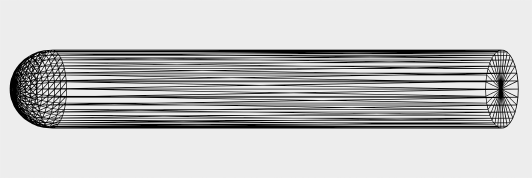}
    \caption{Side view of the borehole component of detector \#1 with perspective rendering; the hemispherical front end of the borehole is visible on the left, where the hemispherical surface is seamlessly connected to the side cylinder and no redundant geometry exists at the hemispherical apex}\label{fig10}
\end{figure}

Figure \ref{fig10} shows the front end of the borehole in the core of detector \#1. For this region, which has a complex hemispherical or arc-shaped bottom structure, DetMesh employs high-precision discretized sampling, fully demonstrating the modeling advantages of triangulated surface geometry.

\subsection{Calibration Results}

To validate the physical accuracy of the simulation calculations, a typical sourceless efficiency calibration scenario was designed in this study. A simulated point source was placed on the detector axis at a distance of 25 cm above the top surface of the detector. To cover the typical energy range of $\gamma$-ray spectrometric measurements and to examine the response of the program to photons of different energies, 13 discrete energy points ranging from 59.5 keV to 1.408 MeV were selected. These energies correspond to the main emission lines of commonly used calibration radionuclides, including $^{\text{241}}$Am, $^{\text{137}}$Cs, $^{\text{60}}$Co, and $^{\text{152}}$Eu, as summarized in Table \ref{tbl3}. During the simulations, both Gadep and the reference Geant4 calculations used $1\times10^7$ particles to ensure that Monte Carlo statistical fluctuations were smaller than the systematic deviations introduced by the geometric and physical models. The full-energy peak efficiencies of the six detectors at the 13 energy points were compared individually.

\begin{table}
\caption{Point-source energies and corresponding radionuclides used in the simulations}\label{tbl3}
\centering
\begin{tabular}{l l}
    \toprule
    \textbf{Nuclide} & \textbf{Photon Energy (keV)} \\
    \midrule
    ${}^{241}\text{Am}$ & 59.54 \\
    ${}^{133}\text{Ba}$ & 81.00, 356.02 \\
    ${}^{152}\text{Eu}$ & 121.78, 244.70, 344.28, 411.13, 443.97, \\
                        & 778.90, 964.06, 1112.09, 1408.02 \\
    ${}^{60}\text{Co}$  & 1332.49 \\
    \bottomrule
\end{tabular}
\end{table}

\begin{figure}
  \centering
   \includegraphics[width=.9\textwidth]{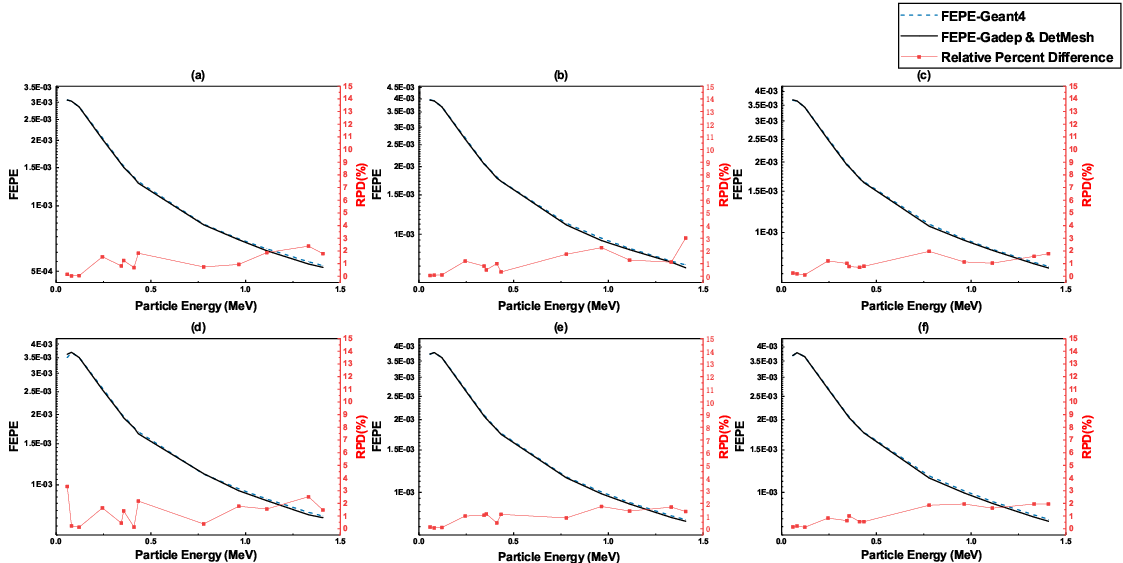}
    \caption{FEPE curves simulated using Gadep–DetMesh and Geant4, together with their differences, for a point source located 250 mm above the top surface of the detector. $1\times10^7$ particles were simulated. In Geant4, the detector was modeled using the built-in modeling strategy and native geometry representation without using triangulated surface models.}\label{fig11}
\end{figure}

Figure \ref{fig11} presents the comparison between the Gadep–DetMesh calculation results and the Geant4 simulation results, together with the distribution of relative deviations. The results show that, among a total of 78 test points (6 detectors $\times$ 13 energies), only two energy points exhibit relative deviations slightly exceeding 3\%, while the deviations of the remaining 76 points are controlled within 3\%. Considering the minor differences between different Monte Carlo programs in the treatment of physical cross-section libraries and the implementation of low-energy electromagnetic physics models, these results demonstrate that the Gadep–DetMesh scheme has high physical consistency and reliability, and can satisfy the requirements of high-precision $\gamma$-ray spectrometric analysis.

\subsection{Acceleration based on GPU}

The CPU benchmark group was based on Geant4 with MPI parallelization enabled. The hardware resources included six computing nodes equipped with Intel Xeon(R) Gold 6148 CPUs, with a total of 60 physical cores used for computation. The GPU experimental group was based on Gadep–DetMesh and was executed on a single NVIDIA GeForce RTX 4090 graphics card. Table \ref{tbl4} lists the detailed computational times for detector \#2 under different particle-history scales, including the total simulation time for all 13 energy points.

\begin{table}
\caption{Computational times of Gadep–DetMesh and Geant4 for monoenergetic point sources located 250 mm above the top surface of the detector}\label{tbl4}
\centering
    \begin{tabular}{ccccc}
        \toprule
        \textbf{Number of Particles} & \textbf{CPU(60 Cores)} & \textbf{GPU (1x 4090)} & \textbf{Speedup} & \textbf{Speedup / Device} \\
        \midrule
        $1 \times 10^7$ & 93 s  & 11.96 s & 7.78$\times$  & $\sim$460$\times$ \\
        $1 \times 10^8$ & 894 s & 66.08 s & 13.53$\times$ & $\sim$800$\times$ \\
        \bottomrule
    \end{tabular}
\end{table}

For the high-statistics simulation with $1\times10^8$ particles, the single-GPU calculation completed all simulations in only 66.08 s, whereas the cluster using 60 high-performance CPU cores still required nearly 15 min. Even under this highly asymmetric comparison between a single GPU card and a CPU cluster, the Gadep–DetMesh scheme achieved a speedup factor of 13.5. It is worth emphasizing that, for ordinary users, building and maintaining a 60-core MPI-parallel computing environment and writing the corresponding Geant4 parallel program involve a very high technical threshold. In contrast, Gadep–DetMesh requires only a personal workstation equipped with a consumer-grade graphics card. This transfer of computing capability from a high-performance cluster to a single desktop workstation represents an orders-of-magnitude improvement in practical efficiency and a substantial reduction in cost for real-world applications.

To further examine the influence of the complex geometric identification logic introduced by DetMesh on GPU performance, a mesh-density sensitivity test was conducted. The discretization parameters of detector \#2 were increased from the default values of $N_\theta = 32$ and $N_\phi = 10$ to $N_\theta = 64$ and $N_\phi = 20$. As a result, the number of generated triangular facets increased from 4416 to 13440. The test results show that, for $1\times10^8$ particles, the GPU computation time increased only slightly from 66.08 s to 68.42 s, corresponding to an increase of only 3.5\%. This result strongly demonstrates that the BVH acceleration structure built into Gadep effectively mitigates the traversal overhead caused by the increased number of triangular facets. Therefore, the proposed framework can maintain excellent computational performance even when handling complex detector models with higher geometric precision.

\section{Discussion}\label{sec4}

Although the results obtained using Gadep–DetMesh are highly consistent with those obtained using Geant4, small systematic differences still exist. These differences can mainly be attributed to two factors. First, different Monte Carlo codes may adopt different interpolation algorithms for low-energy photon cross-section libraries and different treatments of electron multiple-scattering models. Second, the geometry modeling framework in Geant4 is structurally different from the use of triangulated surface models in Gadep. On the other hand, the detector dead layer has a substantial influence on the energy response in the low-energy region. Even minor changes in modeling parameters can significantly affect the detection efficiency of the detector at low energies and lead to relatively large numerical deviations. Nevertheless, the efficiency calibration results obtained using Gadep–DetMesh are in close agreement with those obtained for the same detector model constructed within the Geant4 geometry framework. This strongly demonstrates that the proposed scheme can be reliably applied to sourceless efficiency calibration of detectors.

The improvement in computational performance is not merely a reduction in computation time; more importantly, it lowers the practical threshold for sourceless efficiency calibration. The experimental results show that the computational output of a single consumer-grade graphics card, namely the RTX 4090, is approximately 13.5 times that of a 60-core server cluster. This indicates that sourceless efficiency calibration tasks that previously required high-performance CPU servers can now be completed on personal workstations, or even laptops, equipped with GPUs. This migration of computational capability from centralized high-performance servers to local GPU-equipped workstations makes the proposed scheme highly suitable for deployment in practical settings such as hospital nuclear medicine departments and environmental monitoring stations. It therefore has broad application value.

In addition, the detector triangulated surface modeling framework proposed in this study, DetMesh, is highly lightweight. It can generate detector triangulated surface models with adjustable precision at very low deployment and computational cost. Compared with general-purpose modeling libraries such as CGAL, DetMesh has lower learning, development, and deployment costs. In addition to exporting models, it can also be directly embedded into subsequent computational modules, which provides strong general applicability. DetMesh is also developer-friendly, allowing users and detector manufacturers to conduct further development based on the framework.

At present, the DetMesh framework has been well adapted to standard coaxial HPGe detectors. However, for detectors with non-axisymmetric features, the current parameterized templates still need to be extended. In addition, although the ray-based checking logic resolves the problem of geometric nesting, the branch prediction efficiency of the GPU may decrease when handling extremely complex scenarios involving tens of thousands of triangular facets or more. Future work will focus on developing a more general topological description interface and further optimizing GPU memory-access strategies.

\section{Conclusions}\label{sec5}

To address the time-consuming computation problem in sourceless efficiency calibration of high-purity germanium (HPGe) detectors, this study independently developed DetMesh, a lightweight detector triangulated surface modeling framework. By combining DetMesh with Gadep, a GPU-accelerated Monte Carlo transport program, a complete integrated modeling-and-computation solution was proposed. Through systematic experimental validation and performance evaluation, the following conclusions were obtained.

First, the DetMesh framework enables automated and high-fidelity mapping from detector geometric parameters to three-dimensional triangulated surface models. The framework adopts an object-oriented hierarchical architecture and supports flexible adjustment of the discretization accuracy parameters $\left(N_\theta, N_\phi\right)$. It can accurately describe complex topological structures in coaxial detectors, including the front-edge rounding, the hemispherical bottom of the borehole, and the double-layer nesting structure of the dead layer. This not only resolves the geometric distortion problem associated with conventional voxelized modeling, but also provides customizable precision options for simulation scenarios under different computational-resource constraints.

Second, the Gadep–DetMesh computational scheme demonstrates high physical reliability. Through comparative tests with Geant4 over a broad energy range from 59.5 keV to 1.408 MeV, the relative deviations in full-energy peak efficiency at 78 typical energy points were mostly controlled within 3\%. In particular, the high level of agreement in the low-energy region strongly demonstrates that the parameterized dead-layer modeling strategy proposed in this study can accurately reflect the real energy-response characteristics of the detector.

Third, while maintaining computational accuracy, the proposed scheme achieves an orders-of-magnitude improvement in Monte Carlo simulation speed. The experimental results show that the computational output of a single consumer-grade NVIDIA RTX 4090 graphics card is approximately 13.5 times that of a 60-core Intel Xeon server cluster, demonstrating significant acceleration performance. Combined with the very low deployment threshold of DetMesh, this scheme successfully transfers high-performance computing capability to local workstations, making minute-level, high-precision sourceless efficiency calibration possible on personal workstations. It therefore provides a technically valuable pathway for rapid on-site analysis and large-scale detector calibration in the field of nuclear radiation measurement.

\bibliographystyle{unsrtnat}
\bibliography{references}  

@book{001knoll2010radiation,
  title={Radiation detection and measurement},
  author={Knoll, Glenn F},
  year={2010},
  publisher={John Wiley \& Sons}
}

@article{002eberth2008ge,
  title={From Ge (Li) detectors to gamma-ray tracking arrays--50 years of gamma spectroscopy with germanium detectors},
  author={Eberth, J and Simpson, J},
  journal={Progress in Particle and Nuclear Physics},
  volume={60},
  number={2},
  pages={283--337},
  year={2008},
  publisher={Elsevier}
}

@article{003agostinelli2003geant4,
  title={Geant4—a simulation toolkit},
  author={Agostinelli, Sea and Allison, John and Amako, K al and Apostolakis, John and Araujo, Henrique and Arce, Pedro and Asai, Makoto and Axen, D and Banerjee, Swagato and Barrand, GJNI and others},
  journal={Nuclear instruments and methods in physics research section A: Accelerators, Spectrometers, Detectors and Associated Equipment},
  volume={506},
  number={3},
  pages={250--303},
  year={2003},
  publisher={Elsevier}
}

@article{004allison2016recent,
  title={Recent developments in Geant4},
  author={Allison, John and Amako, Katsuya and Apostolakis, John and Arce, Pedro and Asai, Makoto and Aso, Tsukasa and Bagli, Enrico and Bagulya, A and Banerjee, S and Barrand, GJNI and others},
  journal={Nuclear instruments and methods in physics research section A: Accelerators, Spectrometers, Detectors and Associated Equipment},
  volume={835},
  pages={186--225},
  year={2016},
  publisher={Elsevier}
}

@article{005briesmeister2000mcnptm,
  title={MCNPTM-A general Monte Carlo N-particle transport code},
  author={Briesmeister, Judith F and others},
  journal={Version 4C, LA-13709-M, Los Alamos National Laboratory},
  volume={2},
  year={2000}
}

@article{006moens1981calculation,
  title={Calculation of the absolute peak efficiency of gamma-ray detectors for different counting geometries},
  author={Moens, Luc and De Donder, J and Xi-Lei, Lin and De Corte, Frans and De Wispelaere, Antoine and Simonits, A and Hoste, Julien},
  journal={Nuclear Instruments and Methods in Physics Research},
  volume={187},
  number={2-3},
  pages={451--472},
  year={1981},
  publisher={Elsevier}
}

@article{007sima2002transfer,
  title={Transfer of the efficiency calibration of Germanium gamma-ray detectors using the GESPECOR software},
  author={Sima, O and Arnold, D},
  journal={Applied Radiation and Isotopes},
  volume={56},
  number={1-2},
  pages={71--75},
  year={2002},
  publisher={Elsevier}
}

@article{008hurtado2004monte,
  title={Monte Carlo simulation of the response of a germanium detector for low-level spectrometry measurements using GEANT4},
  author={Hurtado, S and Garc{\i}a-Leon, M and Garc{\i}a-Tenorio, R},
  journal={Applied Radiation and Isotopes},
  volume={61},
  number={2-3},
  pages={139--143},
  year={2004},
  publisher={Elsevier}
}

@article{009boson2008detailed,
  title={A detailed investigation of HPGe detector response for improved Monte Carlo efficiency calculations},
  author={Boson, Jonas and {\AA}gren, G{\"o}ran and Johansson, Lennart},
  journal={Nuclear Instruments and Methods in Physics Research Section A: Accelerators, Spectrometers, Detectors and Associated Equipment},
  volume={587},
  number={2-3},
  pages={304--314},
  year={2008},
  publisher={Elsevier}
}

@article{010budjavs2009optimisation,
  title={Optimisation of the MC-model of a p-type Ge-spectrometer for the purpose of efficiency determination},
  author={Budj{\'a}{\v{s}}, D and Heisel, M and Maneschg, W and Simgen, H},
  journal={Applied Radiation and Isotopes},
  volume={67},
  number={5},
  pages={706--710},
  year={2009},
  publisher={Elsevier}
}

@article{011vidmar2009crystal,
  title={Crystal rounding and the efficiency transfer method in gamma-ray spectrometry},
  author={Vidmar, Tim and Gasparro, Jo{\"e}l},
  journal={Applied Radiation and Isotopes},
  volume={67},
  number={11},
  pages={2057--2061},
  year={2009},
  publisher={Elsevier}
}

@article{012subercaze2022effect,
  title={Effect of the geometrical parameters of an HPGe detector on efficiency calculations using Monte Carlo methods},
  author={Subercaze, Alexandre and Sauzedde, Thibault and Domergue, Christophe and Destouches, Christophe and Philibert, Herve and Fausser, Clement and Thiollay, Nicolas and Gregoire, Gilles and Zoia, Andrea},
  journal={Nuclear Instruments and Methods in Physics Research Section A: Accelerators, Spectrometers, Detectors and Associated Equipment},
  volume={1039},
  pages={167096},
  year={2022},
  publisher={Elsevier}
}

@article{013jia2011gpu,
  title={GPU-based fast Monte Carlo simulation for radiotherapy dose calculation},
  author={Jia, Xun and Gu, Xuejun and Graves, Yan Jiang and Folkerts, Michael and Jiang, Steve B},
  journal={Physics in Medicine \& Biology},
  volume={56},
  number={22},
  pages={7017--7031},
  year={2011}
}

@article{014hissoiny2011gpumcd,
  title={GPUMCD: a new GPU-oriented Monte Carlo dose calculation platform},
  author={Hissoiny, Sami and Ozell, Beno{\^\i}t and Bouchard, Hugo and Despr{\'e}s, Philippe},
  journal={Medical physics},
  volume={38},
  number={2},
  pages={754--764},
  year={2011},
  publisher={Wiley Online Library}
}

@article{015tian2015gpu,
  title={A GPU OpenCL based cross-platform Monte Carlo dose calculation engine (goMC)},
  author={Tian, Zhen and Shi, Feng and Folkerts, Michael and Qin, Nan and Jiang, Steve B and Jia, Xun},
  journal={Physics in Medicine \& Biology},
  volume={60},
  number={19},
  pages={7419--7435},
  year={2015},
  publisher={IOP Publishing}
}

@article{016franciosini2023gpu,
  title={GPU-accelerated Monte Carlo simulation of electron and photon interactions for radiotherapy applications},
  author={Franciosini, Gaia and Battistoni, Giuseppe and Cerqua, Arianna and De Gregorio, Angelica and De Maria, Patrizia and De Simoni, Micol and Dong, Yunsheng and Fischetti, Marta and Marafini, Michela and Mirabelli, Riccardo and others},
  journal={Physics in Medicine \& Biology},
  volume={68},
  number={4},
  pages={044001},
  year={2023},
  publisher={IOP Publishing}
}

@article{ren2026research,
  title={Research on the Advanced GPU-Based Monte Carlo Neutron Transport Methods},
  author={Ren, Yushuo and Li, Zeguang and Lin, Henglong and Wang, Huifu and Wang, Kan},
  journal={Computer Physics Communications},
  pages={110101},
  year={2026},
  publisher={Elsevier}
}

@article{TSOU199430,
  title={Monte Carlo simulation for Compton suppression spectrometer},
  author={Tsou, RH and Lin, Simon C and Kiang, LL},
  journal={Computer physics communications},
  volume={83},
  number={1},
  pages={30--44},
  year={1994},
  publisher={Elsevier}
}

@article{b017goorley2012initial,
  title={Initial MCNP6 release overview},
  author={Goorley, T and James, Michael and Booth, Thomas and Brown, Forrest and Bull, Jeffrey and Cox, Lawrence J and Durkee, Jr and Elson, Jay and Fensin, Michael and Forster, RA and others},
  journal={Nuclear technology},
  volume={180},
  number={3},
  pages={298--315},
  year={2012},
  publisher={Taylor \& Francis}
}

@article{b019garcia2021variance,
  title={Variance-reduction methods for Monte Carlo simulation of radiation transport},
  author={Garc{\'\i}a-Pareja, Salvador and Lallena, Antonio M and Salvat, Francesc},
  journal={Frontiers in Physics},
  volume={9},
  pages={718873},
  year={2021},
  publisher={Frontiers Media SA}
}

@article{b020asano2022photon,
  title={Photon detector response function methodology using MCNP and shift hybrid radiation transport code for wide-area contamination assay applications},
  author={Asano, Ethan and Coleman, D and Davidson, Gregory and Dewji, Shaheen},
  journal={Nuclear Instruments and Methods in Physics Research Section A: Accelerators, Spectrometers, Detectors and Associated Equipment},
  volume={1031},
  pages={166568},
  year={2022},
  publisher={Elsevier}
}

@article{b021zhang2023development,
  title={Development and application of variance reduction technique based on response matrix method in the cosRMC code},
  author={Zhang, Xian and Liu, Shichang and Zhang, Jingyu and Chen, Yixue},
  journal={Annals of Nuclear Energy},
  volume={186},
  pages={109753},
  year={2023},
  publisher={Elsevier}
}

@article{niess2025goupil,
  title={Goupil: A Monte Carlo engine for the backward transport of low-energy gamma-rays},
  author={Niess, Valentin and Vernet, Kinson and Terray, Luca},
  journal={Computer Physics Communications},
  volume={314},
  pages={109653},
  year={2025},
  publisher={Elsevier}
}

@article{soplin2025monte,
  title={Monte Carlo simulation development and implementation of the GiBUU model for neutrino experiments},
  author={Soplin, Leonidas Aliaga and Fern{\'a}ndez, Raquel Castillo and Gustafson, Jasper and Quinn, Declan and Yadav, Shweta},
  journal={Computer Physics Communications},
  volume={311},
  pages={109553},
  year={2025},
  publisher={Elsevier}
}

@article{b022moens1983peak,
  title={Calculation of the peak efficiency of high-purity germanium detectors},
  author={Moens, Luc and Hoste, Julien},
  journal={The International journal of applied radiation and isotopes},
  volume={34},
  number={8},
  pages={1085--1095},
  year={1983},
  publisher={Elsevier}
}

@article{b023wang1995esolan,
  title={HPGe detector absolute-peak-efficiency calibration by using the ESOLAN program},
  author={Wang, Tien-Ko and Mar, Wei-Yang and Ying, Tzung-Hua and Liao, Chi-Hung and Tseng, Chia-Lian},
  journal={Applied radiation and isotopes},
  volume={46},
  number={9},
  pages={933--944},
  year={1995},
  publisher={Elsevier}
}

@article{b024abbas2007direct,
  title={Direct mathematical method for calculating full-energy peak efficiency and coincidence corrections of HPGe detectors for extended sources},
  author={Abbas, Mahmoud I},
  journal={Nuclear Instruments and Methods in Physics Research Section B: Beam Interactions with Materials and Atoms},
  volume={256},
  number={1},
  pages={554--557},
  year={2007},
  publisher={Elsevier}
}

@article{b025jiang1998hybrid,
  title={A hybrid method for calculating absolute peak efficiency of germanium detectors},
  author={Jiang, SH and Liang, JH and Chou, JT and Lin, UT and Yeh, WW},
  journal={Nuclear Instruments and Methods in Physics Research Section A: Accelerators, Spectrometers, Detectors and Associated Equipment},
  volume={413},
  number={2-3},
  pages={281--292},
  year={1998},
  publisher={Elsevier}
}

@article{b026ozben2009hybrid,
  title={A hybrid method to determine efficiency curve of HPGe detectors},
  author={Ozben, Cenap S and Emirhan, Erhan M},
  journal={Applied Radiation and Isotopes},
  volume={67},
  number={6},
  pages={1110--1113},
  year={2009},
  publisher={Elsevier}
}

@article{b027radu2009etna,
  title={ETNA software used for efficiency transfer from a point source to other geometries},
  author={Radu, Daniela and Stanga, Doru and Sima, Octavian},
  journal={Applied Radiation and Isotopes},
  volume={67},
  number={9},
  pages={1686--1690},
  year={2009},
  publisher={Elsevier}
}

@article{b028salman2019hybrid,
  title={Investigation hybrid MCNP/Angle model for calculating the absolute full-energy peak efficiency of HPGe detector},
  author={Salman, A and Ahmed, Z and Allam, Kh A and El-Sharkawy, S},
  journal={Applied Radiation and Isotopes},
  volume={150},
  pages={57--62},
  year={2019},
  publisher={Elsevier}
}

@article{b029bell2012angle,
  title={An investigation of HPGe gamma efficiency calibration software (ANGLE V. 3) for applications in nuclear decommissioning},
  author={Bell, SJ and Judge, SM and Regan, PH},
  journal={Applied Radiation and Isotopes},
  volume={70},
  number={12},
  pages={2737--2741},
  year={2012},
  publisher={Elsevier}
}

@article{b030prozorova2021characterizing,
  title={Characterizing the coaxial HPGe detector using Monte Carlo simulations and evolutionary algorithms},
  author={Prozorova, Irina V and Ghal-Eh, Nima and Bedenko, Sergey V and Popov, Yury A and Prozorov, Andrey A and Vega-Carrillo, Hector R},
  journal={Applied Radiation and Isotopes},
  volume={174},
  pages={109748},
  year={2021},
  publisher={Elsevier}
}

@article{b031lin2023determination,
  title={Determination of detection efficiency on HPGe detector for point-like and volumetric samples based on Geant4 simulations},
  author={Lin, Mu and Wang, Yang and Qin, Zhi},
  journal={Applied radiation and isotopes},
  volume={200},
  pages={110989},
  year={2023},
  publisher={Elsevier}
}

@article{YanGadep,
  author = {Yan, Shuchang and Qiu, Rui and Hu, Ankang and Qu, Shuiyin and Zhou, Yang and Hu, Ziyi and Zhou, Yanhan and Wu, Zhen and Li, Junli},
  title = {Optimized GPU-accelerated Monte Carlo program for real-time dose estimation directly using mesh-type computational phantoms},
  journal = {Nuclear Science and Techniques},
  note = {In Press}
}






\end{document}